\documentclass{iaus}

\usepackage{graphicx}
\usepackage{natbib}

\title[~~EMU] 
{Evolutionary Map of the Universe}

\author[Norris]   
{Ray P. Norris$^{1\star}$}

\affiliation{$^1$CSIRO Astronomy and Space Science, PO Box 76, Epping, NSW, 1710, Australia\\
$\star$email: {\tt Ray.Norris@csiro.au}
}

\pubyear{2011} 
\volume{284}  
\pagerange{1--12}
\jname{The Spectral Energy Distribution of Galaxies}
\editors{R.J. Tuffs \&  C.C.Popescu, eds.}

\def\deg {^{\circ} }
\def\sqdeg {\,deg$^2$}

\def\ujybm {\,$\mu$Jy/beam}

\def \etal {\rm ~{\it \etal},~}

\def\apj {{\it Ap.~J.}}

\def\aj {{\it A.~J.}}

\def\mnras {{\it MNRAS}}

\def\pasa {{\it PASA}}

\begin{document}

\maketitle

\begin{abstract}
  EMU is a wide-field radio continuum survey planned for the new Australian Square Kilometre Array Pathfinder (ASKAP) telescope, due to be completed in 2012. The primary goal of EMU is to make a deep ($\sim$10$\mu$Jy/bm rms) radio continuum survey of the entire Southern Sky at 1.4 GHz, extending as far North as +30$\deg$ declination, with a 10 arcsec resolution. EMU is expected to detect and catalog about 70 million galaxies, including typical star-forming galaxies up to z=1, powerful starbursts to even greater redshifts, and AGNs to the edge of the Universe. EMU will undoubtedly discover new classes of object. Here I present the science goals and survey parameters.
  
\end{abstract}

\firstsection 
\section{Introduction}

Until recently, most large radio surveys could only detect radio-loud Active Galactic Nuclei (AGN) and 
very nearby star-forming galaxies. As a result, radio data were of little interest to those modelling the Spectral Energy Distribution (SED) of populations of galaxies. However, we are now going through a period of massive change in radio-astronomy. New technology being developed for the pathfinder instruments of the Square Kilometre Array (SKA) are enabling radio surveys far deeper than before, so that radio data will start to become increasingly important for studies of multiwavelength SEDs.
Here I describe the largest such survey, EMU (Evolutionary Map of the Universe), which will reach a similar sensitivity ($\sim$ 10 \ujybm) to the deepest current small-area surveys, but over the entire visible sky. At that sensitivity, EMU will be able to trace the evolution of galaxies over most of the lifetime of the Universe.

\begin{figure}[h]
\begin{center}
\includegraphics[scale=0.4, angle=0]{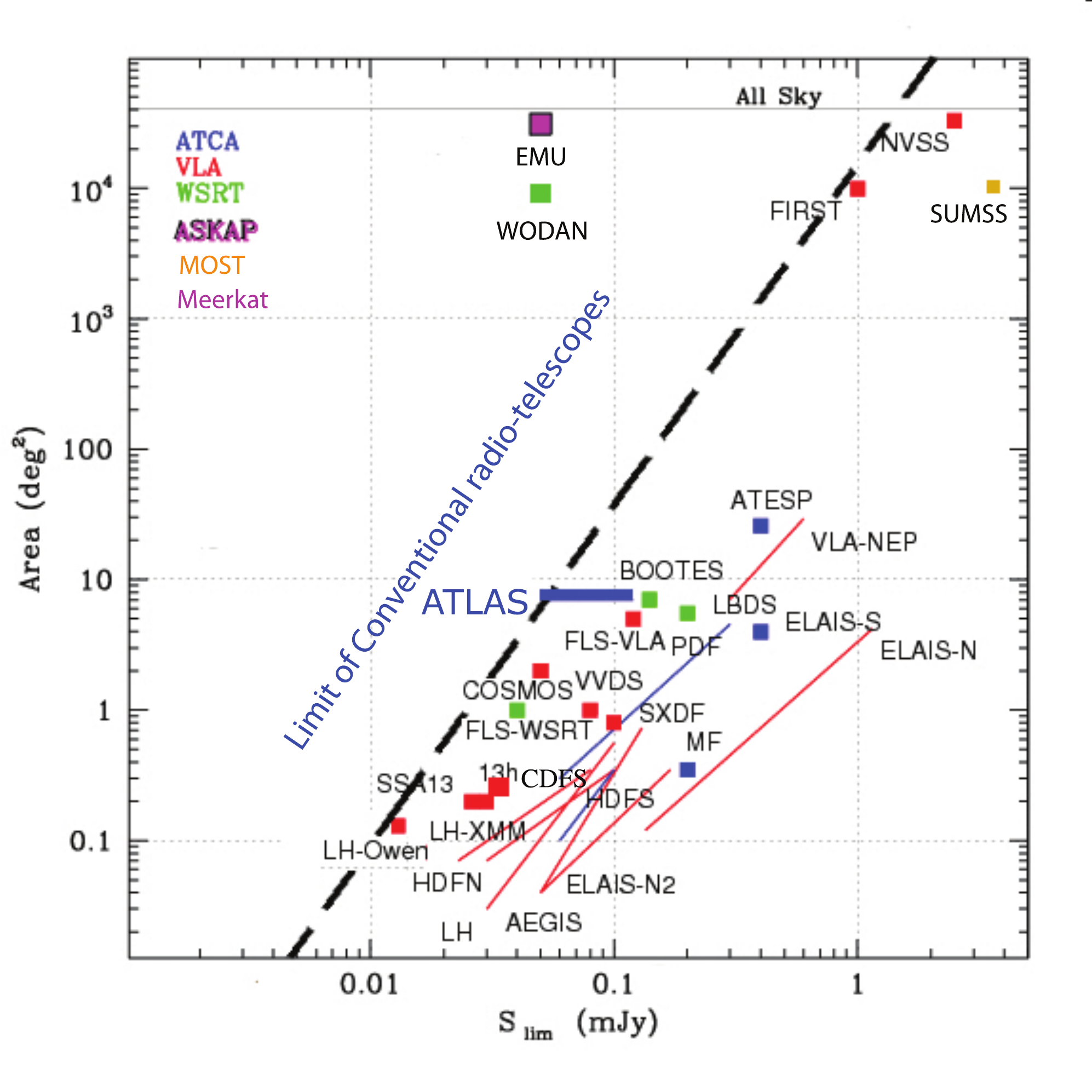}
\caption{Comparison of EMU with existing deep 20 cm radio surveys. Horizontal axis is 5-$\sigma$ sensitivity, and vertical axis shows the sky coverage. The diagonal dashed line shows the approximate envelope of existing surveys, which is largely determined by the availability of telescope time. The squares in the top-left represent the EMU survey, discussed in this paper, and the complementary WODAN \citep{Rottgering10b} survey which has been proposed to cover the sky North of $+30\deg$. 
}
\label{fig1}
\end{center}
\end{figure}

 Fig. 1 shows the major 20-cm continuum radio surveys. The largest existing radio survey, shown in the top right, is the wide but shallow NRAO VLA Sky Survey (NVSS) \citep{Condon98}. The most sensitive existing radio survey is the deep but narrow Lockman Hole observation  \citep{Owen08} in the lower left. All current surveys are bounded by a diagonal line that roughly marks the limit of available telescope time of current-generation radio telescopes. The region to the left of this line is currently unexplored, and this area of observational phase space presumably contains as many potential new discoveries as the region to the right.

\section {ASKAP: the Australian SKA Pathfinder}
The Australian SKA Pathfinder (ASKAP) is a new radio telescope being built both to test and develop aspects of potential SKA science and technology, and to demonstrate the capabilities of the Australian SKA candidate site. However, ASKAP is a major telescope in its own right, likely to generate significant new astronomical discoveries. ASKAP  \citep{Johnston07, Johnston08, Deboer09} will consist of 36 12-metre antennas spread over a region 6 km in diameter. Although the array of antennas is no larger than many existing radio telescopes, the feed array at the focus of each antenna is revolutionary, with a phased-array feed (PAF) of 96 dual-polarisation pixels, designed to work in a frequency band of 700--1800 MHz, with an instantaneous bandwidth of 300 MHz. This will replace the single-pixel feeds that are almost universal in current-generation synthesis radio telescopes. As a result, ASKAP will have a  field of view up to 30 \sqdeg, enabling it to survey the sky thirty times faster than existing synthesis arrays, and allowing surveys of a scope that cannot be contemplated with current-generation telescopes. To ensure good calibration, the antennas are a novel 3-axis design, with the feed and reflector rotating  to mimic  the effect of an equatorial mount,  ensuring a constant position angle of the PAF and sidelobes on the sky. 

As the result of a competitive selection process, ASKAP design is being driven by ten survey science projects, with priority being given to two key projects: WALLABY (an all-sky HI survey) and EMU (an all-sky radio continuum survey).

\section{EMU: Evolutionary Map of the Universe}
The primary goal of EMU \citep{Norris11a} is to make a deep (10 \ujybm\ rms) radio continuum survey of the entire Southern Sky, extending as far North as $+30\deg$. EMU will cover roughly the same fraction (75\%) of the sky as the benchmark NVSS survey \citep{Condon98}, but will be 45 times more sensitive, and will have an angular resolution (10 arcsec) 4.5 times better. Because of the excellent short-spacing \emph{uv} coverage of ASKAP, EMU will also have higher sensitivity to extended structures. 
Like most radio surveys, EMU will adopt a 5-$\sigma$ cutoff, leading to a source detection threshold of 50 \ujybm. EMU is expected to generate a catalogue of about 
70 million galaxies, and all
radio data from the EMU survey will be placed in the public domain as soon as the data quality has been assured.

EMU differs from many previous surveys in that  the survey includes cross-identification with major surveys at other wavelengths, and will produce public-domain VO-accessible catalogues as ``value-added'' data products. This is facilitated by the growth in the number of large southern hemisphere telescopes and associated planned major surveys spanning all wavelengths. In addition, EMU is collaborating closely with complementary radio surveys such as WODAN  \citep{Rottgering10b}, LOFAR  \citep{Rottgering10a}, and Meerkat-MIGHTEE. To help plan EMU, a pilot survey, ATLAS, is being conducted over 7 sq. deg. of the CDFS-SWIRE and ELAIS-SWIRE fields \citep{Norris06, Middelberg08}.

%

\begin{figure}[h]
\begin{center}
\includegraphics[scale=0.2, angle=0]{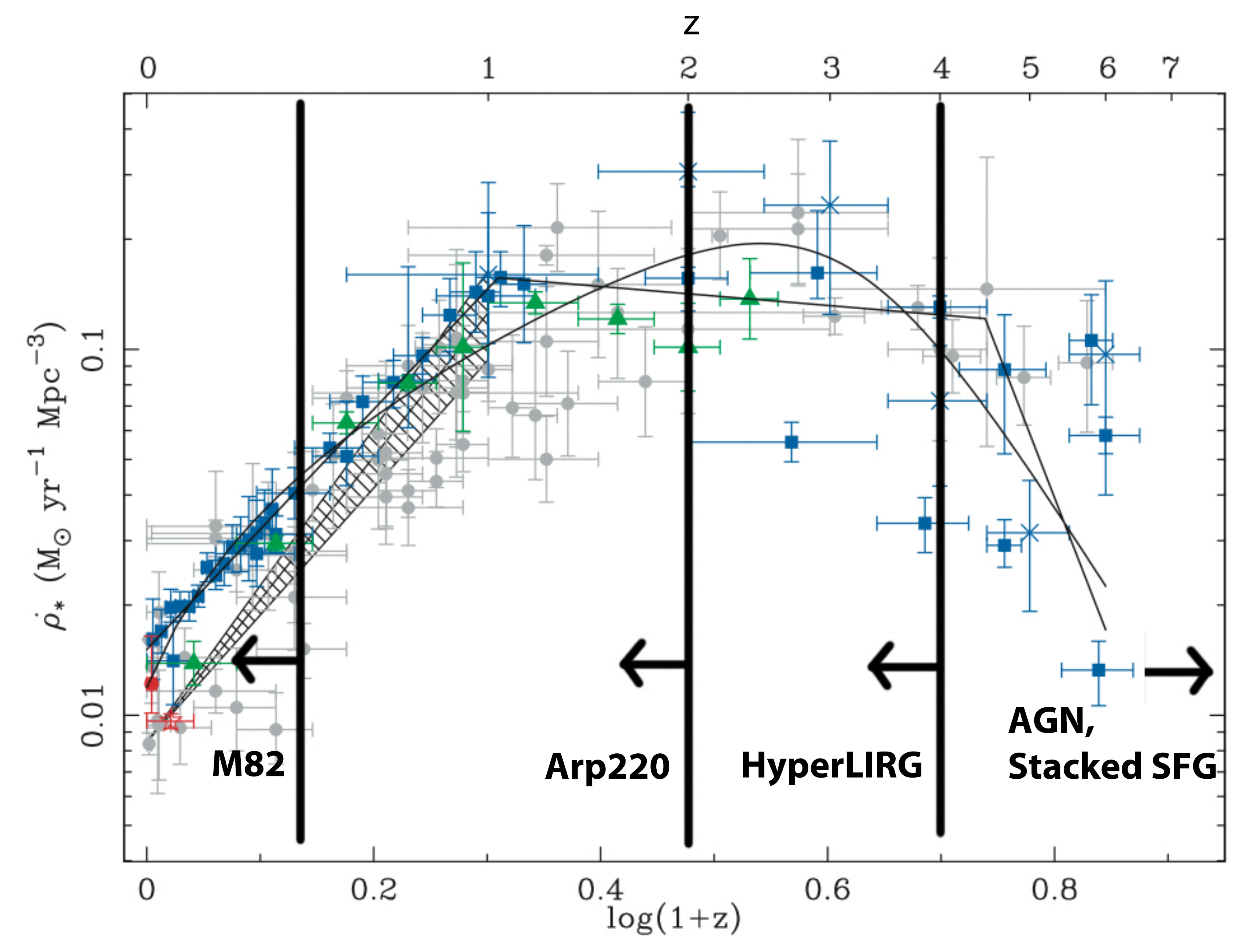}
\caption{The cosmic star formation rate as a function of cosmic time, adapted from \cite{Hopkins06}, showing the EMU 5$\sigma$ detection limits for an M82, Arp220, and a HyperLIRG. AGNs and stacked star-forming galaxies can be detected to arbitrarily high redshifts. The plot is very uncertain beyond z=2, and may be level to a higher redshift, because of uncertain extinction corrections. EMU should be able to distinguish betwen alternative models at high redshift. 
}
\label{fig2}
\end{center}
\end{figure}

\section{Science}

Broadly, the key science goals for EMU, detailed in \cite{Norris11a}, are:
\begin{itemize}
\item To trace the evolution of star-forming galaxies from $z=2$ to the present day, and stack to even higher redshift. Radio data are unaffected by dust, enabling us to use radio flux as a sensitive and accurate measure of star formation rate
\citep{Garn09}.

\begin{figure}[h]
\begin{center}
\includegraphics[scale=0.3, angle=0]{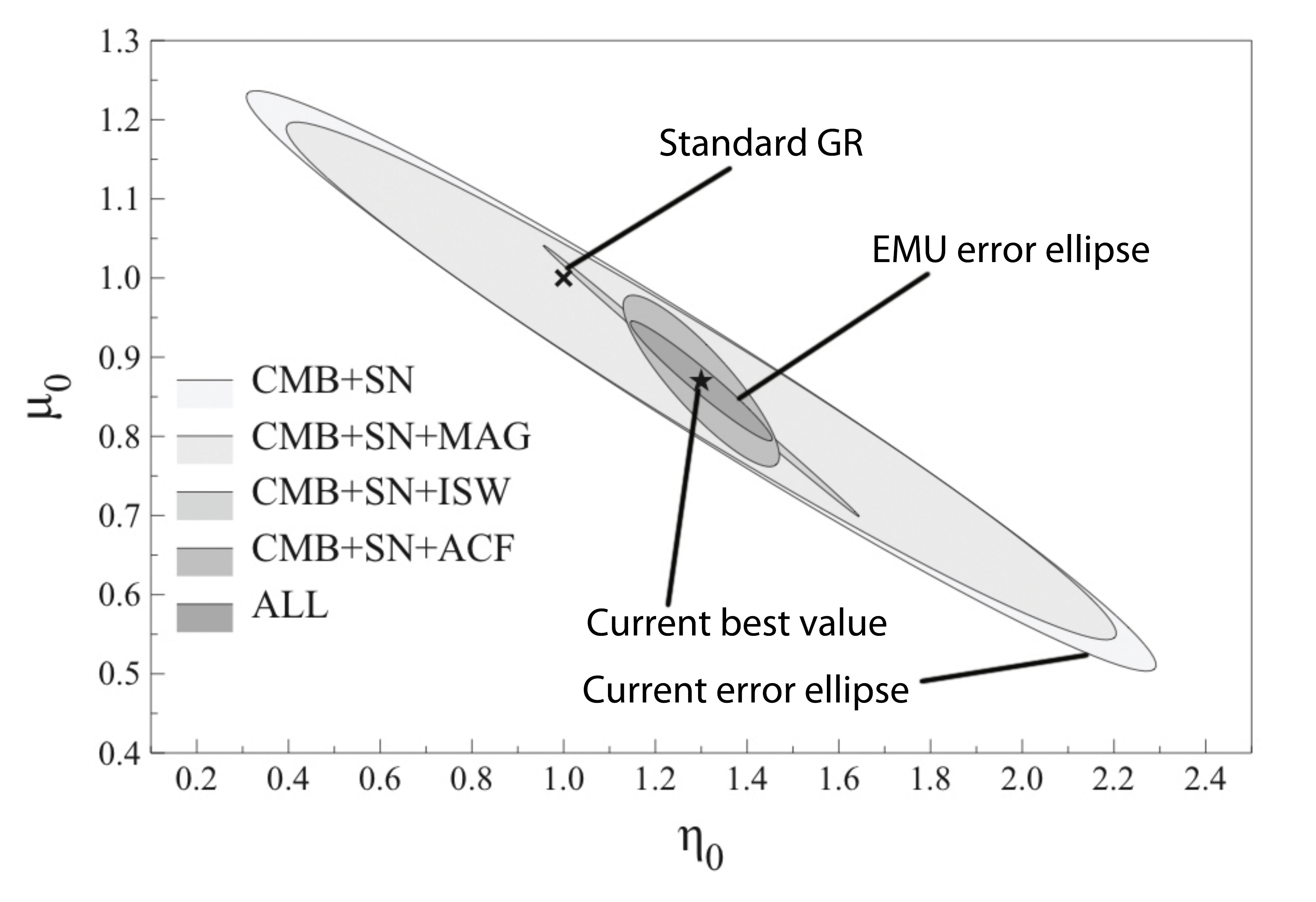}
\caption{How EMU will constrain the parameters of modified gravity, adapted from \cite{Raccanelli11}. The much smaller error ellipse provided by EMU may exclude alternatives to General Relativity such as braneworld models.
}
\label{fig3}
\end{center}
\end{figure}

\item To trace the evolution of massive black holes throughout the history of the Universe, and understand their relationship to star formation. Particularly interesting are those AGN  buried beneath many magnitudes of extinction, which are invisible at optical/IR wavelengths \citep[e.g.][]{Norris11b, Norris11c}.
\item To use the distribution of radio sources to explore the large-scale structure and cosmological parameters of the Universe, and to test fundamental physics. For example, by using brute-force statistics to measure the effect of weak gravitational lensing and the Integrated Sachs-Wolfe effect, \cite{Raccanelli11} have shown that EMU will make the best measurement yet of dark energy and modified gravity parameters (see Fig. 3).
\item To determine how radio sources populate dark matter halos, as a step towards understanding the underlying astrophysics of clusters and halos. For example, we expect to detect thousands, and possibly hundreds of thousands, of clusters by using wide-angle-tail galaxies as cluster probes \citep[e.g.][]{Mao10}.
\item To create the most sensitive wide-field atlas of Galactic continuum emission yet made in the Southern Hemisphere, addressing areas such as star formation, supernovae, and Galactic structure. For example, we expect to detect thousands of new radio stars, and have started the SCORPIO pilot survey (Umana et al., in preparation) of 4 sq. deg. of the Galactic Plane, to guide our survey strategy.
\item To explore an uncharted region of observational parameter space, with a high likelihood of finding new classes of object. In particular, we will systematically mine the database of EMU data to search for sources whose properties lie outside the parameters of known objects or phenomena.
\end{itemize}

\end{document}